\documentclass{article}
\usepackage[utf8]{inputenc}
\usepackage{array}
\usepackage{tabularx}
\usepackage[
style=phys
]{biblatex} 
\usepackage{graphicx}
\usepackage{authblk}
\usepackage{siunitx}
\usepackage{accsupp}

\title{Simple coplanar waveguide resonator mask targeting metal-substrate interface}
\author[1]{Cameron J. Kopas}
\author[1]{Ella Lachman}
\author[2,3,4,*]{Corey Rae H. McRae}
\author[1]{Yuvraj Mohan}
\author[1]{Josh Y. Mutus}
\author[1]{Ani Nersisyan}
\author[1]{Amrit Poudel}
\affil[1]{Rigetti Computing, 775 Heinz Ave, Berkeley CA 94701}
\date{\today}
\affil[2]{
Department of Physics, University of Colorado, Boulder, Colorado 80309, USA}
\affil[3]{
Department of Electrical, Computing and Energy Engineering, University of Colorado, Boulder, Colorado 80309, USA}
\affil[4]{ 
National Institute of Standards and Technology, Boulder, Colorado 80305, USA
}%
\affil[*]{ 
coreyrae.mcrae@colorado.edu
}%

\begin{document}

\maketitle

\begin{abstract}
    This white paper presents a single-layer mask, found in Ref.~\cite{Mask}\footnote{https://github.com/Boulder-Cryogenic-Quantum-Testbed/simple-resonator-mask}, designed for fabrication of superconducting microwave resonators towards 1:1 comparisons of dielectric losses from the metal-substrate interface. Finite-element electromagnetic simulations are used to determine participation ratios of the four major regions of the on-chip devices, as well as to confirm lack of crosstalk between neighboring devices and demonstrate coupling tunability over three orders of magnitude. This mask is intended as an open-source community resource for facilitating precise and accurate comparisons of materials in the single-photon, millikelvin regime.
\end{abstract}

\section{Introduction}

Superconducting microwave resonators are a critical tool for materials loss characterization in superconducting quantum computing \cite{mcrae2020materials}. Resonator measurements allow us to distinguish between sources of loss impacting device performance as well as to compare losses from a variety of materials growth and fabrication processes including surface treatments and interface engineering methods. Coplanar waveguide (CPW) resonators are commonly implemented in multiqubit circuits for qubit control, readout, and qubit-qubit coupling. As they are generally patterned from the same single superconducting thin film that makes up the qubit capacitor pads, they are ideal for measuring losses of materials and interfaces that dominate qubit relaxation.

At power levels near a single photon of average power and temperatures in the tens of mK, superconducting CPW resonators are sensitive to materials losses due to spurious two-level states (TLS) in on-chip amorphous dielectrics, crystal defects and impurities, and contaminants within regions in the device such as the substrate, the metal-substrate (MS) interface, the substrate-air (SA) interface, and the metal-air (MA) interface \cite{Muller2019}.


The total loss induced in a resonator is equivalent to the inverse of the internal quality factor, $Q_i = 1/\delta$. Further, the loss angle induced in a resonator due to TLS can be broken down as:
\begin{equation}
    \delta_{TLS} = \sum_j {p_j \delta_j},
    \label{loss_sum}
\end{equation}
where $p_j$ is the participation ratio of the $j$-th device region, and $\delta_j$ is the TLS loss of the $j$-th region. Assuming each region of interest is thin enough that the electric field changes only across the surface, the participation ratio for region $j$ of surface area $S$ and thickness $t_j$ can be further expressed in terms of the electric field as:
\begin{equation}
    p_j \approx t_j \frac{\int_S \vec{E}^{(j)} \cdot \epsilon_j \vec{E}^{*(j)} dS}{U_{tot}},
\end{equation}
where $\vec{E}^{(j)}$ is the electric field in this region, $\epsilon_j$ is its relative permittivity, and $U_{tot}$ is the total energy density. This can be solved using finite-element full-wave driven electromagnetic simulations. The level of sensitivity to regions in the device can be tuned to an extent by modifying device geometry. Changes in the CPW gap width $g$, conductor width $s$, and trench depth and geometry determine TLS participation ratios. 

The gold standard for dielectric loss measurement of individual materials within a superconducting quantum device is outlined in Refs.~\cite{Woods2019} and \cite{Melville2020}. This method allows the identification of loss for each individual interface in the device as well as the bulk dielectric, at the cost of requiring the measurement of ~120 devices across four geometrically extreme CPW designs. For the majority of materials experiments, this precision is both unnecessary and experimentally infeasible, and the knowledge of comparative performance is sufficient. For these types of experiments, it is beneficial to have a simple mask set with many devices per chip and a high sensitivity to the device region of interest, while maintaining ease of fabrication.

For experiments varying the MS interface, such as those focused on substrate cleaning \cite{Nersisyan2019,Wisbey2010,Bruno2015,Earnest2018}, resonators with a strong sensitivity to this interface allow higher-precision loss comparisons, and may even allow the identification of performance-improving techniques that would otherwise be obscured. In this white paper, we present a mask set designed to have an increased participation ratio of the MS interface as described in Ref.\cite{Melville2020}, while avoiding confounding experimental factors by maintaining simplicity in device design and associated fabrication procedures. This single-layer mask provides other features such as $\sim$50 $\Omega$ matching on both silicon and sapphire substrates, eight resonators per chip to lend itself to statistical analysis, and tapered bond pads to reduce reflections. We also provide participation values for each device region, as well as simulations of coupler arm lengths $\ell_c$ to reach coupling quality factors ($Q_c$) spanning three orders of magnitude.

We hope this open-source mask\cite{Mask} will be adopted as a standard across the burgeoning field of materials investigations for superconducting quantum computing. This design is intended to facilitate interlaboratory 1:1 comparisons between materials sets, opening the door for a higher level of device optimization than is currently possible in the field.

\section{Chip design}

\begin{figure}
    \centering
    \includegraphics[width=0.9\textwidth]{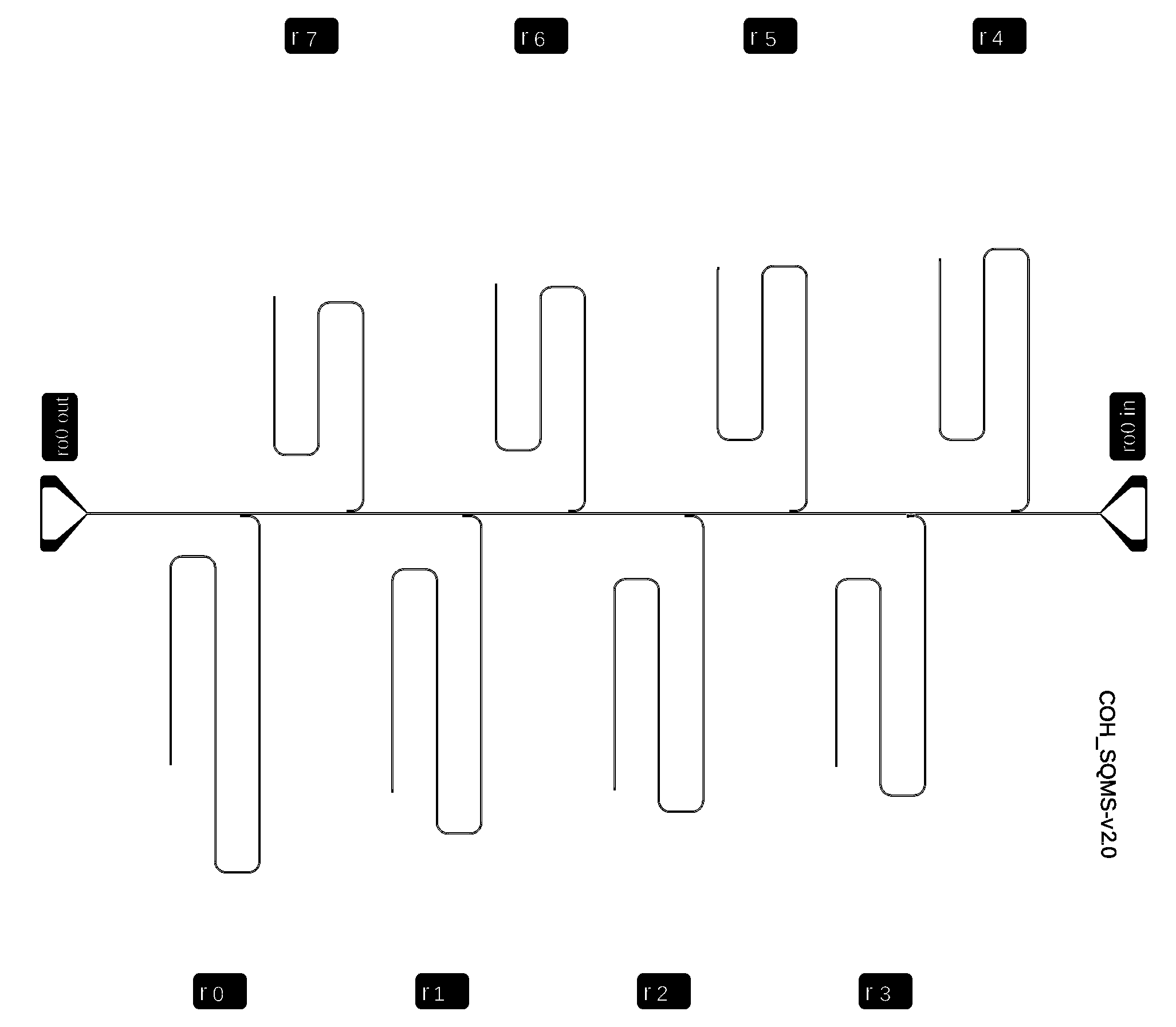}
    \caption{Two-port single-layer chip design with overall chip dimensions of 7.5$\,$mm x 7.5$\,$mm. Eight multiplexed quarter wave resonators are coupled to a single common feedline, and tapered, 50$\,\Omega$-matched bond pads can be seen at the left and right of the chip.}
    \label{fig:chip_design}
\end{figure}

This simple mask is a single-layer 7.5$\,$mm x 7.5$\,$mm chip design composed of eight quarter wave CPW resonators inductively coupled to a single feedline (Fig. \ref{fig:chip_design}). The feedline is connected to wirebonding pads on either side of the chip. Each chip contains resonator, feedline, and design name labels. Dicing alignment marks are not included. Individual resonator parameters from simulation are included in Table \ref{tab:specs}.

\begin{figure}
    \centering
    \includegraphics[width=0.5 \textwidth]{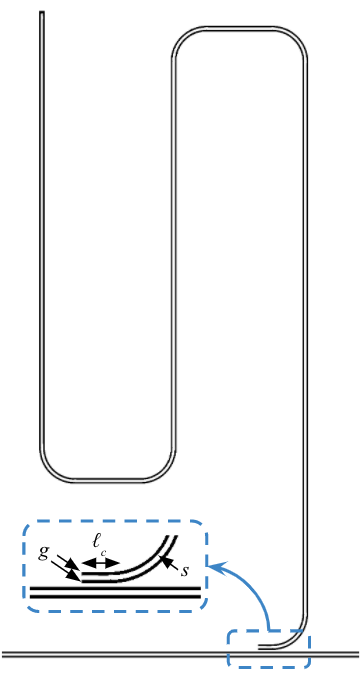}
    \caption{Graphic of a single resonator with gap width $g$ and conductor width $s$ coupled to the feedline by an inductive coupling arm of length $\ell_c$. Areas where the metallic layer is removed are shown in black, and metallized areas are shown in white.}
    \label{fig:resonator}
\end{figure}

A higher-resolution image of a single quarter wave resonator is shown in Fig. \ref{fig:resonator}. The top end of the resonator far from the feedline is open, whereas the end close to the feedline is shorted to ground, forming a voltage node. This induces inductance-dominated coupling between the resonator and the feedline. Both ends of the resonator are filleted to remove field singularities and corresponding losses that could arise from sharp corners. 

\begin{center}
\begin{table}
\caption{\label{tab:specs}High-level chip design specifications.}
\centering
\begin{tabular}{cc}
\hline
\hline
Chip dimensions & $\SI{7.5}{\milli\metre}$ x $\SI{7.5}{\milli\metre}$ \\
\hline
CPW trace width $s$ & $\SI{6}{\micro\metre}$ \\
 \hline
CPW gap width $g$ & $\SI{3}{\micro\metre}$ \\
\hline
Feedline impedance & $\SI{50}{\ohm}$ \\
\hline
Resonators per chip & 10 \\
\hline
Frequency range & 4.6 - 7.4 GHz \\
\hline
Substrate $\epsilon_{r}$ & 11.45 \\
\hline
Smallest feature size & $\SI{3}{\micro\metre}$ \\
\hline
\hline
\end{tabular}
\end{table}
\end{center}

\begin{center}
\begin{table}
\caption{\label{tab:sims}Simulated resonator parameters.}
\centering
\begin{tabular}{cccc}
\hline
\hline
 Resonator &
Simulated &
Simulated &
Simulated \\
label & $f_0$ & decay rate & $Q_c$
\\
 & [MHz] & [MHz]&
\\
 \hline
 \hline
 \\
\hline
 r0 &
4600 &
0.0094 & 
495k 
\\
\hline
 r1 &
5000 &
0.0102 &
496k
\\
\hline
 r2 & 
5400 &
0.0110 &
497k
\\
\hline
 r3 &
5800 &
0.0118 &
499k
\\
\hline
 r4 &
6200 &
0.0126 &
497k
\\
\hline
 r5 &
6600 &  
0.0134 &
498k
\\
\hline
 r6 &
7000 & 
0.0142 &
499k
\\
\hline
 r7 &
7400 &
0.0146 &
512k
\\
\hline
\hline
\end{tabular}
\end{table}
\end{center}

\section{Device simulation}

Table~\ref{tab:sims} summarizes resonator parameters determined by full-wave finite element electromagnetic simulations. Resonance frequencies $f_0$ are expected to vary from simulation by up to 50 MHz due to variation in propagation speed. Here, the characteristic impedance $Z_0$ of the resonator is $\sim$ $\SI{50}{\ohm}$ assuming a silicon substrate ($\epsilon_{r} = 11.45$), but will not deviate significantly from $\SI{50}{\ohm}$ for the implementation of a sapphire substrate.

\begin{figure}
    \centering
    \includegraphics[width=0.9\textwidth]{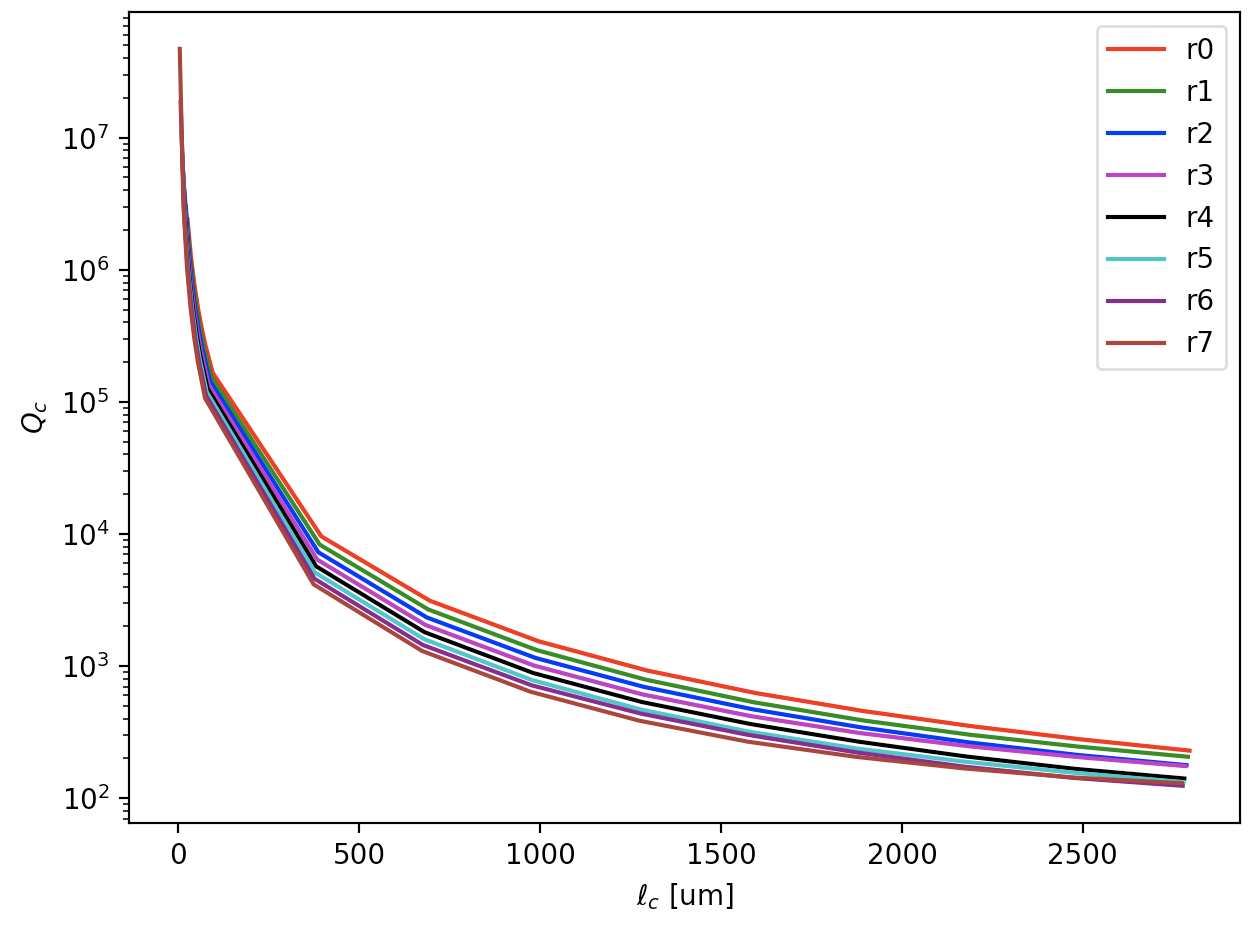}
    \caption{Resonator coupling quality factor $Q_c$ as a function of coupler length $\ell_c$ for resonators r0 to r7. $Q_c$ is tunable by more than three orders of magnitude upon varying the length of the resonator arm parallel to the feedline (the coupler). Color curves correspond to different resonators on the chip.}
    \label{fig:qc_vs_length}
\end{figure}

The inductive coupling strength is controlled by the length of the arm parallel to the feedline (the coupler). In this mask~\cite{Mask}, the coupler is designed such that $Q_c \sim$ 500,000. Since critical coupling is optimal for accurate fitting of $Q_i$ from the Lorentzian resonator transmission model \cite{mcrae2020materials}, this mask is well-designed for accurate measurements of materials which induce a resonator performance of $Q_i \sim 500,000$. Materials systems with higher or lower loss can be accurately measured by lengthening or shortening the coupling arm. Fig.~\ref{fig:qc_vs_length} shows $Q_c$ as a function of coupler length for all eight designed resonators, where the coupler length is defined as the portion of the resonator arm that runs parallel to the feedline. Here, $Q_c$ decreases by more than three orders of magnitude as the coupler length increases.

\begin{figure}
    \centering
    \includegraphics[width=0.6 \textwidth]{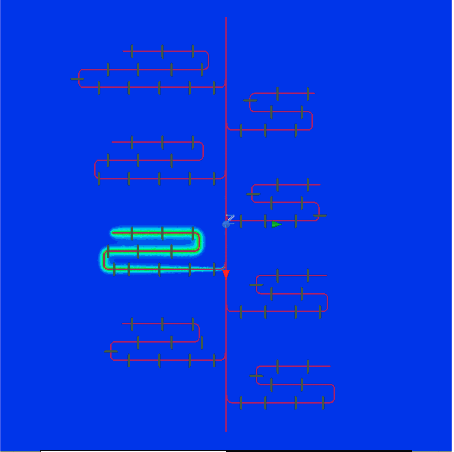}
    \caption{Eigenmode of one of the resonators. Modes are strongly localized around resonators with minimal leakage.}
    \label{fig:eigenmodes}
\end{figure}

Fig.~\ref{fig:eigenmodes} demonstrates that resonator eigenmodes are strongly localized in this design, with minimal field leaking into the neighboring resonators or the feedline, so crosstalk is not anticipated to be a significant loss channel.

\section{Participation of interfaces}

\begin{figure}
    \centering
    \includegraphics[width=0.9\textwidth]{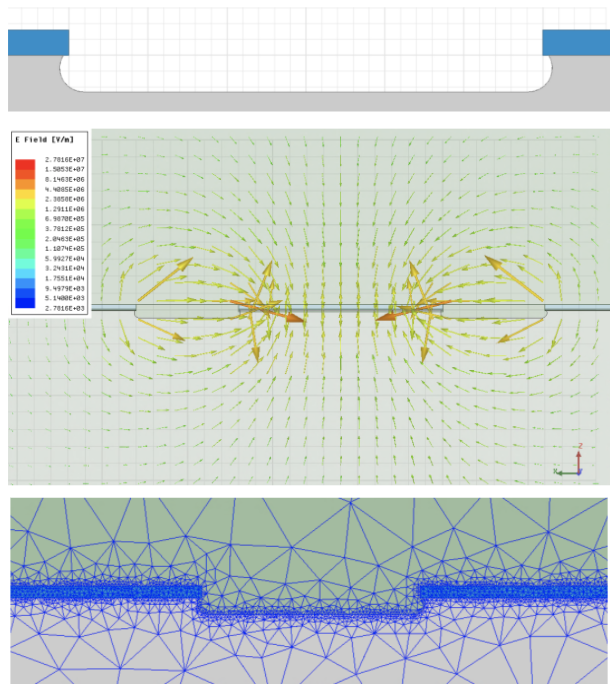}
    \caption{Details of the finite-element electromagnetic simulation of the participation of regions within a $\SI{3}{\micro\metre}$ gap CPW resonator with 100 nm of isotropic trenching. Top: Cross-section of simulated gap region, with metal shown as blue and substrate shown as grey. Middle: Electric field diagram. Bottom: Mesh density.}
    \label{fig:HFSS}
\end{figure}

The participation ratios of the MS, SA, MA, and substrate regions of a $\SI{3}{\micro\metre}$ gap CPW resonator with 100 nm of isotropic trenching (as shown in Fig.~\ref{fig:HFSS}) are reported in Table~\ref{tab:participation}. Assuming interface and surface losses are around three orders of magnitude higher than substrate loss as in Ref.~\cite{Woods2019}, the devices on this mask will be dominated by interface losses, with the MS interface having the largest participation by a factor of two.

\begin{center}
\begin{table}
\caption{\label{tab:participation}Participation of resonator regions determined by simulation of a $\SI{3}{\micro\metre}$ gap CPW resonator with 100 nm of isotropic trenching.}
\centering
\begin{tabular}{ccccc}
\hline
\hline
 $p_{MS}$ &
$p_{SA}$ &
$p_{MA}$ &
$p_{substrate}$ &
$C$ (fF/um)\\
 \hline
0.26 &
0.13 &
0.008 &
89 &
0.147
\\
\hline
\hline
\end{tabular}
\end{table}
\end{center}

\section{Wirebonding}

\begin{figure}
    \centering
    \includegraphics[width=0.85 \textwidth]{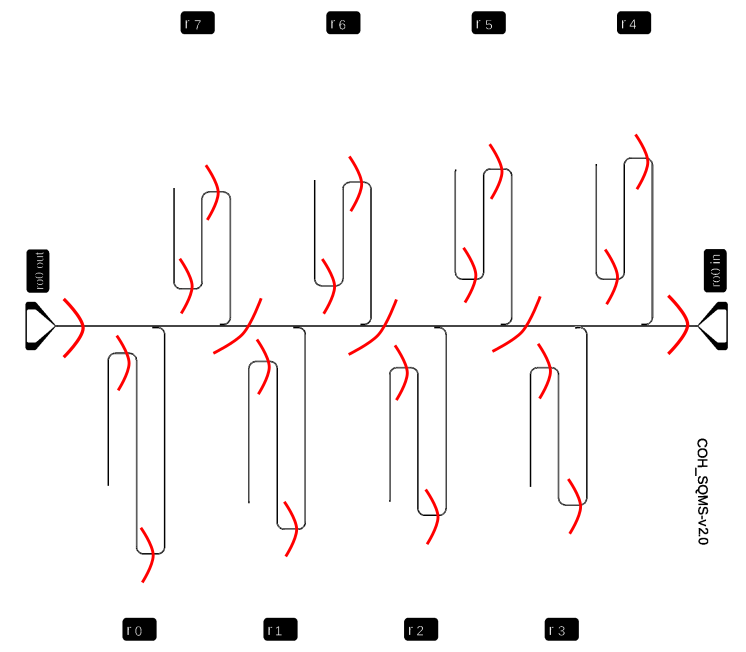}
    \caption{Recommended internal wirebonding schematic, with isolation bonds over the resonators and feedline marked in red.}
    \label{fig:wirebonds}
\end{figure}

A suggested map for wirebond location and density in presented in Fig.~\ref{fig:wirebonds}. Wide resonators meanders allow wirebonding between meanders, connecting otherwise disconnected ground planes and suppressing slotline modes~\cite{Wenner2011}.

\section{Acknowledgements}

This material is based upon work supported by the U.S. Department of Energy, Office of Science, National Quantum Information Science Research Centers, Superconducting Quantum Materials and Systems Center (SQMS) under the Contract No. DE-AC02-07CH11359.

\printbibliography

\end{document}